\documentclass[12pt]{article}
\usepackage{amsmath}
\usepackage{graphicx}
\usepackage{bm}
\usepackage{fancyhdr}
\usepackage{amssymb}
\begin{document}



\def\a{\alpha}
\def\b{\beta}
\def\d{\delta}
\def\e{\epsilon}
\def\g{\gamma}
\def\h{\mathfrak{h}}
\def\k{\kappa}
\def\l{\lambda}
\def\o{\omega}
\def\p{\wp}
\def\r{\rho}
\def\t{\tau}
\def\s{\sigma}
\def\z{\zeta}
\def\x{\xi}
\def\V={{{\bf\rm{V}}}}
 \def\A{{\cal{A}}}
 \def\B{{\cal{B}}}
 \def\C{{\cal{C}}}
 \def\D{{\cal{D}}}
\def\K{{\cal{K}}}
\def\O{\Omega}
\def\R{\bar{R}}
\def\T{{\cal{T}}}
\def\L{\Lambda}
\def\f{E_{\tau,\eta}(sl_2)}
\def\E{E_{\tau,\eta}(sl_n)}
\def\Zb{\mathbb{Z}}
\def\Cb{\mathbb{C}}

\def\R{\overline{R}}

\def\beq{\begin{equation}}
\def\eeq{\end{equation}}
\def\bea{\begin{eqnarray}}
\def\eea{\end{eqnarray}}
\def\ba{\begin{array}}
\def\ea{\end{array}}
\def\no{\nonumber}
\def\le{\langle}
\def\re{\rangle}
\def\lt{\left}
\def\rt{\right}

\newtheorem{Theorem}{Theorem}
\newtheorem{Definition}{Definition}
\newtheorem{Proposition}{Proposition}
\newtheorem{Lemma}{Lemma}
\newtheorem{Corollary}{Corollary}
\newcommand{\proof}[1]{{\bf Proof. }
        #1\begin{flushright}$\Box$\end{flushright}}

\baselineskip=20pt

\newfont{\elevenmib}{cmmib10 scaled\magstep1}
\newcommand{\preprint}{
   \begin{flushleft}
   \end{flushleft}\vspace{-1.3cm}
   \begin{flushright}\normalsize
   \end{flushright}}
\newcommand{\Title}[1]{{\baselineskip=26pt
   \begin{center} \Large \bf #1 \\ \ \\ \end{center}}}
\newcommand{\Author}{\begin{center}
   \large \bf
Fakai Wen${}^{a,b}$, Tao Yang${}^{a,b}$\footnote{Corresponding author:
yangt@nwu.edu.cn}, Zhanying Yang${}^{b,c}$,
~Junpeng Cao${}^{d,e,f}$,~ Kun Hao${}^{a,b}$ and Wen-Li
Yang${}^{a,b,g}\footnote{Corresponding author: wlyang@nwu.edu.cn}$
 \end{center}}
\newcommand{\Address}{\begin{center}

     ${}^a$Institute of Modern Physics, Northwest University,
     Xi'an 710069, China\\
     ${}^b$Shaanxi Key Laboratory for Theoretical Physics Frontiers,
Xi'an 710069, China\\
     ${}^c$School of Physics, Northwest University, Xi'an 710069, China\\
     ${}^d$Institute of Physics, Chinese
Academy of Sciences, Beijing 100190, China\\
     ${}^e$School of Physical Sciences, University of Chinese Academy of
Sciences, Beijing, China\\
     ${}^f$Collaborative Innovation Center of Quantum Matter, Beijing,
     China\\
     ${}^g$Beijing Center for Mathematics and Information Interdisciplinary Sciences, Beijing, 100048,  China

   \end{center}}

\preprint \thispagestyle{empty}
\bigskip\bigskip\bigskip

\Title{Thermodynamic limit and boundary energy of the $su(3)$ spin
chain with non-diagonal boundary fields} \Author

\Address \vspace{1cm}

\begin{abstract}
We investigate the thermodynamic limit of the $su(n)$-invariant  spin
chain models with unparallel boundary fields. It is found that the
contribution of the inhomogeneous term in the associated $T-Q$ relation to the
ground state energy does vanish in the thermodynamic limit. This fact allows us to calculate
the boundary energy of the system. Taking the $su(2)$ (or the XXX) spin chain and the $su(3)$ spin chain as concrete examples,
we have studied the corresponding boundary energies of the  models.  The method used in this paper can
be generalized to study the thermodynamic properties and boundary
energy of other high rank models with non-diagonal boundary fields.

\vspace{1truecm} \noindent {\it PACS:} 75.10.Pq, 02.30.Ik, 71.10.Pm


\end{abstract}

\newpage



\section{Introduction}
\label{intro} \setcounter{equation}{0}

Exactly solvable models have played essential roles in many areas of
physics, such as ultracold atoms \cite{RevModPhys.85.1633},
condensed matter physics
\cite{CambridgeUniversityPress.9780511524332, arXiv.0008018}, the
AdS/CFT correspondence \cite{IntJTheorPhys.38.1113,
LettersMathematicalPhysics.99.1}, equilibrium and non-equilibrium
statistical physics \cite{baxter1982exactly, PhysRevLett.95.240601,
JPhA.40.R333, J.Stat.Mech.1411.P11032, PhysRevA.90.062301,
tome2015stochastic}. The thermodynamic properties of these models,
for example, the specific heat, susceptibility and elementary
excitations, which can be obtained by using the thermodynamic Bethe
ansatz (TBA) \cite{JMP.10.1115}, have attracted a great attention
due to the analytical results can be compared with experimental data
directly \cite{ RevModPhys.85.1633,
CambridgeUniversityPress.9780511524332, JMP.10.1115,
PhysRevLett.19.1312, ComputersinPhysics.12.565, PhysRevB.73.212413}.

In the frame of the off-diagonal Bethe ansatz (ODBA) method
developed recently \cite{wang2015off}, a large class of integrable
models without $U(1)$ symmetry, thus lack of obvious reference
state, can be solved exactly, which attracts general interest, such
as the spin-$1/2$ chain with arbitrary boundary fields
\cite{NuclearPhysicsB.875.152, NuclearPhysicsB.877.152, SIGMA.9.072,
J.Stat.Mech.1502.P02001, NuPhB.899.229}, the $su(n)$ spin chain with
non-diagonal boundary fields
\cite{JournalofHighEnergyPhysics.04.143}, the high spin Heisenberg
chain \cite{JPhysA.46.442002}, the one-dimensional Hubbard model
with arbitrary boundary magnetic fields
\cite{NuclearPhysicsB.879.98}, the XYZ spin chain with odd site
number \cite{NuclearPhysicsB.886.185}, the spin-$1/2$ torus
\cite{PhysRevLett.111.137201} and the Izergin-Korepin Model with
generic open boundaries \cite{JHEP.06.128}. Naturally, the
thermodynamic limit of those models becomes a subject of intense
research \cite{arXiv.1309.6456, JPhysA.47.032001, NuclPhysB.884.17}.
However, the corresponding Bethe Ansatz equations (BAEs) obtained by
using the ODBA method have much complicated structure due to the
inhomogeneous term in the $T$-$Q$ relation, which makes the direct
employment of the TBA method to approach the thermodynamic limit of
those models very involved \cite{wang2015off, arXiv.1309.6456,
JPhysA.47.032001}.

Nevertheless, some important progresses have been made recently
\cite{arXiv.1309.6456, JPhysA.47.032001, NuclPhysB.884.17}. For the
spin-$1/2$ isotropic quantum spin chain with arbitrary
boundary fields, the pioneering work of Jiang et al. \cite{arXiv.1309.6456}
showed that the two boundaries are decoupled from each other in the
thermodynamic limit. In addition, Nepomechie
et al. presented the thermodynamic limit and boundary (or
surface) energy of the model with an expansion up to the second order
in terms of small non-diagonal boundary terms
\cite{JPhysA.47.032001}. For the open XXZ spin chain with generic
boundary fields, a method to address this problem was proposed in
Refs. \cite{wang2015off, NuclPhysB.884.17} based on the fact that
the system has some degenerate points, at which the BAEs become the
usual production ones and can be studied by the TBA. In the
thermodynamic limit, these degenerate points become dense thus
we can use the properties at these degenerate points to approach the
real thermodynamics of the systems. The thermodynamic limit and surface energy were
calculated for arbitrary imaginary crossing parameter
\cite{NuclPhysB.884.17} and for real crossing parameter with a
constraint \cite{wang2015off}, and then applied to the study of the quantum impurities \cite{And16}. However, the results about the
model with arbitrary boundary fields are still missing. Therefore, to
obtain the thermodynamic limit and boundary energy of the models
solved by the ODBA is still an open question and is worth further
study.

In this paper, we study the thermodynamic limit of the $su(n)$-invariant  spin
chain models with unparallel boundary fields by taking the XXX spin-$1/2$ chain
and the $su(3)$-invariant chain with unparallel boundary fields \cite{wang2015off,
NuclearPhysicsB.875.152} as concrete examples. Because of the difficulties arising
from the inhomogeneous term in the $T$-$Q$ relation, the
first thing should be addressed is the contribution of the
inhomogeneous term. Through the analysis of the
finite-lattice systems, it is found  that the contribution of the
inhomogeneous term to the ground state energy {\it does} reduce to zero when
the size of the system tends to infinity. Namely, the inhomogeneous term in
the $T-Q$ relation can be ignored in the thermodynamic limit. We note that, even though  without the inhomogeneous
term, the $T-Q$ relation still contains the non-diagonal boundary fields, whose contribution can be identified by
calculating the boundary energy of the model. Comparison of the boundary energy from
the analytic expressions with that from the Hamiltonian by the
extrapolation method shows that they coincide with each other
very well. This further demonstrates that the neglected inhomogeneous term
does not affect the physical properties of the studied system in the thermodynamic limit.

The paper is organized as follows. Section 2 serves as an introduction to our notations for
the inhomogeneous $su(n)$-invariant spin chains with generic boundary fields.
In Section 3, we focus on the $su(2)$-invariant (or the XXX spin-$1/2$) open spin chain
with the most general non-diagonal boundary terms. With the help of the Bethe ansatz solution for the finite size system,
we study the thermodynamic limit and boundary energy of the model. The results for the $su(3)$-invariant case
are given in Section 4. We summarize our results and give
some discussions in Section 5.

\section{$su(n)$-invariant spin chain with generic boundary fields}
\setcounter{equation}{0}
Let ${\rm\bf V}$ denote a $n$-dimensional linear space with an orthonormal basis $\{|i\rangle|i=1,\cdots,n\}$, which endows the fundamental representation of
$su(n)$ algebra. The $su(n)$-invariant $R$-matrix $R(u)\in {\rm End}({\rm\bf V}\otimes
{\rm\bf V})$  is given by \cite{3-sun,3-sun1}
\begin{eqnarray}
&&R_{12}(u)=u+\eta P_{1,2},\label{R-matrix-1}
\end{eqnarray}
where $u$ is the spectral parameter and $\eta$ is
the crossing parameter (without losing the generality we set $\eta=1$ in the following part of this paper).
The $R$-matrix satisfies the quantum Yang-Baxter equation (QYBE)
\begin{eqnarray}
 R_{12}(u_1-u_2)R_{13}(u_1-u_3)R_{23}(u_2-u_3)=
 R_{23}(u_2-u_3)R_{13}(u_1-u_3)R_{12}(u_1-u_2), \label{QYB}
\end{eqnarray}
and possesses the following properties:
\begin{eqnarray}
 &&\hspace{-1.45cm}\mbox{
 Initial condition}:\hspace{42.5mm}R_{12}(0)=  P_{1,2},\label{Initial}\\
 &&\hspace{-1.5cm}\mbox{
 Unitarity}:\hspace{28.5mm}R_{12}(u)R_{21}(-u)= \rho_1(u)\,{\rm id},\quad \rho_1(u)=-(u+1)(u-1),\label{Unitarity}\\
 &&\hspace{-1.5cm}\mbox{
 Crossing-unitarity}:\quad
 R^{t_1}_{12}(u)R_{21}^{t_1}(-u-n)
 =\rho_2(u)\,\mbox{id},\quad \rho_2(u)=-u(u+n),
 \label{crosing-unitarity}\\
 &&\hspace{-1.4cm}\mbox{Fusion conditions}:\hspace{22.5mm}\, R_{12}(-1)=-2 P^{(-)}_{1,2}, \quad
 R_{12}(1)=2 P^{(+)}_{1,2}.\label{Fusion}
\end{eqnarray}
Here $R_{21}(u)=P_{1,2}R_{12}(u)P_{1,2}$,
$P^{(\mp)}_{1,2}=\frac{1}{2}\{1\mp P_{1,2}\}$ is anti-symmetric
(symmetric) project operator in the tensor product space  ${\rm\bf
V} \otimes {\rm\bf V} $, and $t_i$ denotes the transposition in the
$i$-th space. Here and below we adopt the standard notation: for any
matrix $A\in {\rm End}({\rm\bf V})$, $A_j$ is an embedding operator
in the tensor space ${\rm\bf V}\otimes {\rm\bf V}\otimes\cdots$,
which acts as $A$ on the $j$-th space and as an identity on the
other factor spaces; $R_{ij}(u)$ is an embedding operator of
$R$-matrix in the tensor space, which acts as an identity on the
factor spaces except for the $i$-th and $j$-th ones.

Let us introduce the ``row-to-row"  (or one-row ) monodromy matrix
$T(u)$, which is an $n\times n$ matrix with operator-valued elements
acting on ${\rm\bf V}^{\otimes N}$, \bea T_0(u)
=R_{0N}(u)R_{0\,N-1}(u)\cdots
R_{01}(u).\label{2Mon-V-1} \eea The QYBE implies that one-row monodromy matrix $T(u)$ satisfies the
Yang-Baxter relation
\begin{eqnarray}
R_{00^\prime}(u-v)T_{0}(u)T_{0^\prime}(v)
=T_{0^\prime}(v)T_{0}(u)R_{00^\prime}(u-v).\label{RTT}
\end{eqnarray}
Integrable open chain can be constructed as follows
\cite{Alc87,Skl88}. Let us introduce a pair of $K$-matrices $K^-(u)$
and $K^+(u)$. The former satisfies the reflection equation (RE) \cite{Che84,Skl88}
\bea &&R_{12}(u_1-u_2)K^-_1(u_1)R_{21}(u_1+u_2)K^-_2(u_2)\no\\
 &&~~=
K^-_2(u_2)R_{12}(u_1+u_2)K^-_1(u_1)R_{21}(u_1-u_2),\label{RE-V}
\eea and the latter  satisfies the dual RE
\bea
&&R_{12}(u_2-u_1)K^+_1(u_1)R_{21}(-u_1-u_2-n)K^+_2(u_2)\no\\
&&~~~~~~= K^+_2(u_2)R_{12}(-u_1-u_2-n)K^+_1(u_1)R_{21}(u_2-u_1).
\label{DRE-V}\eea
For open spin-chains, instead of the standard
``row-to-row" monodromy matrix $T(u)$ (\ref{2Mon-V-1}), one needs to
consider  the
 ``double-row" monodromy matrix ${\cal{J}}(u)$
\bea
  {\cal{J}}_0(u)&=&T_0(u)K_0^-(u)\hat{T}_0(u),  \label{Mon-V-0}\\
  \hat{T}_0(u)&=&R_{01}(u)R_{02}(u)\ldots
  R_{0N}(u).\no
\eea Then the double-row transfer matrix $t(u)$ of the open spin
chain is given by \bea
t(u)=tr_0\{K^+_0(u){\cal{J}}_0(u)\}.\label{trans} \eea From the QYBE
and the (dual) RE, one may check that the transfer matrices with
different spectral parameters commute with each other:
$[t(u),t(v)]=0$. Thus $t(u)$ serves as the generating functional of
the conserved quantities, which ensures the integrability of the
system described by the Hamiltonian
\begin{eqnarray}
&&H=\frac{\partial \ln t(u)}{\partial
u}|_{u=0} \nonumber \\
&&\quad = 2\sum_{j=1}^{N-1}P_{j,j+1} +  \frac{tr_0
{K^+_0}^\prime(0)}{tr_0 K^+_0(0)} +2 \frac{tr_0 K_{0}^+(0)
P_{0N}}{tr_0 K^+_0(0)} + \{K^{-}_1(0)\}^{-1}{K_{1}^-}^\prime(0).
\label{oh}
\end{eqnarray}
The commutativity of the transfer matrices with different spectral
parameters implies that they have  common eigenstates. Let
$|\Psi\rangle$ be a common eigenstate of $t(u)$, which does
not depend upon $u$, with the eigenvalue $\Lambda(u)$,
\bea
t(u)|\Psi\rangle=\Lambda(u)|\Psi\rangle.\label{Eigenvalue-open}
\eea
Using the ODBA method \cite{wang2015off,JournalofHighEnergyPhysics.04.143}, the corresponding
eigenvalue $\Lambda(u)$ is given in terms of an inhomogeneous $T-Q$ relation
\cite{PhysRevLett.111.137201,NuclearPhysicsB.875.152}. The aim of this paper is to
investigate the thermodynamic limit ($N\longrightarrow\infty$) of the model by the exact solutions
obtained in \cite{NuclearPhysicsB.875.152, JournalofHighEnergyPhysics.04.143}.

\section{Boundary energy of the XXX spin-$1/2$ chain with arbitrary boundary fields}
\setcounter{equation}{0}

The Hamiltonian given by (\ref{oh}) of the XXX spin-$1/2$ chain with unparallel
boundary fields reads \cite{wang2015off, NuclearPhysicsB.875.152}
\begin{eqnarray}\label{Hamiltonian}
  H=\sum_{j=1}^{N-1} (\vec \sigma_{j}\cdot \vec
  \sigma_{j+1})+\frac{1}{p}\sigma_{1}^{z}+\frac{1}{q}(\sigma_{N}^{z}+\xi\sigma_{N}^{x})+N,
\end{eqnarray}
where $\vec \sigma_{j}$ is the Pauli matrix at site $j$, and $p, q$ and
$\xi$ are all arbitrary real boundary parameters to ensure a
hermitian Hamiltonian. The corresponding $K$-matrices $K^{\pm}(u)$ are given by
\begin{eqnarray}
K^-(u)=\left(\begin{array}{cc}
p+u & 0\\
0 & p-u
\end{array}\right),\label{K-2}
\end{eqnarray}
and
\begin{eqnarray}
K^+(u)=\left(\begin{array}{cc}
q+u+1 & \xi(u+1)\\
\xi(u+1) & q-u-1
\end{array}\right). \label{K+2}
\end{eqnarray}
The eigenvalue $\Lambda(u)$ of the corresponding transfer matrix is given in terms of the  inhomogeneous $T-Q$
relation \cite{NuclearPhysicsB.875.152}
\begin{eqnarray}
    \Lambda(u)&=&\frac{2(u+1)^{2N+1}}{2u+1}(u+p)\left[(1+\xi^{2})^{\frac{1}{2}}u+q\right]\frac{Q(u-1)}{Q(u)} \nonumber \\[6pt]
&&+\frac{2u^{2N+1}}{2u+1}(u-p+1)\left[(1+\xi^{2})^{\frac{1}{2}}(u+1)-q\right]\frac{Q(u+1)}{Q(u)} \nonumber \\[6pt]
&&+2\left[1-(1+\xi)^{\frac{1}{2}}\right]\frac{[u(u+1)]^{2N+1}}{Q(u)},
\label{T_Q_relation}
\end{eqnarray}
where the function $Q(u)$ can be parameterized as
\begin{eqnarray}\label{Q_function}
  Q(u)=\prod_{j=1}^{N}(u-\lambda_{j})(u+\lambda_{j}+1),
\end{eqnarray}
and the $N$ Bethe roots $\{\lambda_{j}|j=1,\cdots, N\}$ should
satisfy a set of BAEs,
\begin{eqnarray}
  &&\left(\frac{\lambda_{j}+1}{\lambda_{j}}\right)^{2N+1}\frac{(\lambda_{j}+p)
  \left[(1+\xi^{2})^{\frac{1}{2}}\lambda_{j}+q\right]}{(\lambda_{j}-p+1)\left[(1+\xi^{2})^{\frac{1}{2}}(\lambda_{j}+1)-q\right]}
  =\nonumber \\[6pt]
  &&-\frac{\left[1-(1+\xi^{2})^{\frac{1}{2}}\right](2\lambda_{j}+1)(\lambda_{j}+1)^{2N+1}}
  {(\lambda_{j}-p+1)\left[(1+\xi^{2})^{\frac{1}{2}}(\lambda_{j}+1)-q\right]\prod_{l=1}^{N}(\lambda_{j}-\lambda_{l}-1)(\lambda_{j}+\lambda_{l})}\nonumber \\[6pt]
  &&-\prod_{l=1}^{N}\frac{(\lambda_{j}-\lambda_{l}+1)(\lambda_{j}+\lambda_{l}+2)}{(\lambda_{j}-\lambda_{l}-1)(\lambda_{j}+\lambda_{l})},
  \quad j=1,\cdots, N.\label{BAEs}
\end{eqnarray}
The eigenvalue of the Hamiltonian is given in terms of the Bethe roots by
\begin{eqnarray}\label{Energy}
  E=\sum_{j=1}^{N}\frac{2}{\lambda_{j}(\lambda_{j}+1)}+2N-1+\frac{1}{p}+\frac{(1+\xi^{2})^{\frac{1}{2}}}{q}.
\end{eqnarray}

\subsection{Contribution of the inhomogeneous term to the ground state energy}
In order to study the contribution of the inhomogeneous term (the last term in Eq.(\ref{T_Q_relation})) to the ground state energy,
we first consider the $T-Q$
relation without the inhomogeneous term\footnote{It should be emphasized that, for a finite $N$, $\Lambda_{hom}(u)$
is different from the exact eigenvalue $\Lambda(u)$ given by (\ref{T_Q_relation}).}, i.e.,
\begin{eqnarray}\label{T_Q_relation01a}
  \Lambda_{hom}(u)&=&\frac{2(u+1)^{2N+1}}{2u+1}(u+p)\left[(1+\xi^{2})^{\frac{1}{2}}u+q\right]\frac{Q(u-1)}{Q(u)} \nonumber \\[6pt]
&&+\frac{2u^{2N+1}}{2u+1}(u-p+1)\left[(1+\xi^{2})^{\frac{1}{2}}(u+1)-q\right]\frac{Q(u+1)}{Q(u)}.
\end{eqnarray}
The singular property of the $T-Q$ relation (\ref{T_Q_relation01a}) gives the
following BAEs
\begin{eqnarray}\label{BAE4}
  \left(\frac{\mu_{j}-\frac{i}{2}}{\mu_{j}+\frac{i}{2}}\right)^{2N}\frac{(\mu_{j}-i\bar{p})
  \left(\mu_{j}-i\bar{q}\right)}{(\mu_{j}+i\bar{p})\left(\mu_{j}+i\bar{q}\right)}
  = \prod_{l\neq
  j}^{M}\frac{(\mu_{j}-\mu_{l}-i)(\mu_{j}+\mu_{l}-i)}{(\mu_{j}-\mu_{l}+i)(\mu_{j}+\mu_{l}+i)},
\end{eqnarray}
where we have put $\lambda=i\mu-\frac{1}{2}$,
$\bar{p}=p-\frac{1}{2}$ and
$\bar{q}=q(1+\xi^{2})^{-\frac{1}{2}}-\frac{1}{2}$.

We define the contribution of the inhomogeneous term to the ground state energy as
\begin{eqnarray}\label{DeltaE}
  E_{inh}=E_{hom}-E_{true}.
\end{eqnarray}
Here $E_{hom}$ is the ground state energy of the XXX spin-$1/2$ chain calculated by the homogeneous $T-Q$ relation (\ref{T_Q_relation01a}).
In this case (i.e., without the inhomogeneous term), the number of Bethe roots reduces to $M=N/2$, when an even $N$ is assumed.
 Then energy
$E_{hom}$ is given by equation (\ref{Energy}) with the constraint
(\ref{BAE4}). $E_{true}$ is the ground state energy of the
Hamiltonian (\ref{Hamiltonian}), which can be obtained by either using the density matrix renormalization group (DMRG)
\cite{JSMTE.05.05001} or solving the BAEs
(\ref{BAEs}) directly. We have checked that the ground state
energy $E_{true}$ obtained by these two methods are the same.
\begin{figure}[ht]
  \centering
  \includegraphics[width=5.5in]{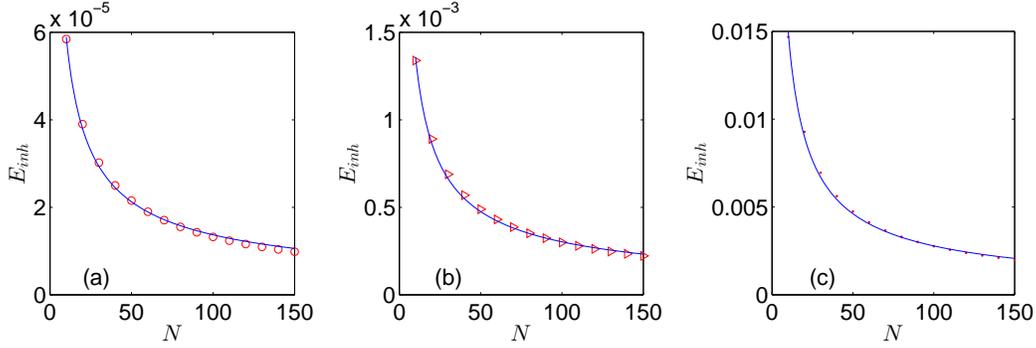}\\
  \caption{The contribution of the inhomogeneous term to the ground state energy
$E_{inh}$ versus the system size $N$. The data can be fitted as
$E_{inh}=\gamma_{1} N^{\beta_{1}}$. Due to the fact $\beta_{1}<0$,
when the $N$ tends to infinity, the contribution of the
inhomogeneous term tends to zero. Here $p=8$, $q=4$, (a)
$\xi=\frac{1}{8}$, $\gamma_{1}=0.000253$ and $\beta_{1}=-0.6334$;
  (b) $\xi=\frac{5}{8}$, $\gamma_{1}=0.006096$ and $\beta_{1}=-0.6521$; (c) $\xi=\frac{25}{8}$,
  $\gamma_{1}=0.080180$ and $\beta_{1}=-0.7297$.}\label{SU2XXX-DEvsN}
\end{figure}

From the fitted curves in Figure \ref{SU2XXX-DEvsN}, we find the power law relation
between $E_{inh}$ and $N$, i.e., $E_{inh}=\gamma_{1} N^{\beta_{1}}$.
Due to the fact that $\beta_{1}<0$, the value of $E_{inh}$ tends to zero when
the size of the system tends to infinity, which means that the
inhomogeneous term in Eq.(\ref{T_Q_relation}) can be neglected in the
thermodynamic limit.

\subsection{Boundary energy}

Now, the boundary energy is ready to be calculated. We consider the case of
$\bar{p}, \bar{q}\geq 0$, in which all the Bethe roots are real at the ground state. Taking the logarithm of equation (\ref{BAE4}), we obtain
\begin{eqnarray}
  &&2\arctan\left(\frac{2\mu_{j}}{2\bar{p}}\right)+2\arctan\left(\frac{2\mu_{j}}{2\bar{q}}\right)+4N\arctan\left(2\mu_{j}\right)
  =2\pi I_{j} \nonumber \\[6pt]
  &&+\sum_{l=1}^{M}\left[2\arctan\left(\frac{2(\mu_{j}-\mu_{l})}{2}\right)
  +2\arctan\left(\frac{2(\mu_{j}+\mu_{l})}{2}\right)\right]-2\arctan\left(2\mu_{j}\right),
\label{BAEs5}
\end{eqnarray}
where $I_{j}$ is a set of quantum numbers. If we define
\begin{eqnarray}
   Z(\mu_{j})=\frac{I_{j}}{2N}.
\end{eqnarray}
It turns to be a continuous function in the thermodynamic limit as the distribution of Bethe roots is
continuous, i.e., $Z(\mu_{j}) \to Z(u)$. Taking the derivative of
$Z(u)$ with respect to $u$, we obtain the density of states as
\begin{eqnarray}\label{rho}
  \rho(u)=a_{1}(u)+\frac{1}{2N}\left[a_{2\bar{p}}(u)+a_{2\bar{q}}(u)+a_{1}(u)
  -\delta(u)\right]-\int_{-\infty}^{\infty}a_{2}(u-v)\rho(v)d v,
\end{eqnarray}
where
\begin{eqnarray}
  a_{n}(u)=\frac{1}{2\pi}\frac{n}{u^2+\frac{n^2}{4}}.
\end{eqnarray}
The energy density of the ground state is
\begin{eqnarray}
    e_{g}&=&-2\pi \int_{-\infty}^{\infty}a_{1}(\mu)\rho(\mu)d\mu+1-\frac{1}{2N}+\frac{1}{2Np}+\frac{(1+\xi^{2})^{\frac{1}{2}}}{2Nq} \nonumber \\[6pt]
    &=&1-2\ln(2)+{O}(N^{-1}).
\end{eqnarray}

The boundary energy is given by \cite{PhysRevA.4.386,
JournalofPhysicsA.20.5677, JournalofPhysicsA.23.761}
\begin{eqnarray}
    E_{b}(p,q,\xi)&=&\lim_{N\rightarrow\infty}\left[E_{0}(N;p,q,\xi)-2E_{0}^{periodic}(N)\right]\nonumber \\[6pt]
    &=&-4\pi N\int_{-\infty}^{\infty}\tilde{a}_{1}(\omega)\delta\tilde{\rho}(\omega)d\omega-1+\frac{1}{p}+\frac{(1+\xi^{2})^{\frac{1}{2}}}{q}.
\end{eqnarray}
The density deviation from that of the periodic case satisfies
\begin{eqnarray}
  \delta\rho(u)=\frac{1}{2N}\left[a_{2\bar{p}}(u)+a_{2\bar{q}}(u)+a_{1}(u)
  -\delta(u)\right]-\int_{-\infty}^{\infty}a_{2}(u-v)\delta\rho(v)d v.
\end{eqnarray}
With the help of the Fourier transformation, we have
\begin{eqnarray}
  \delta\tilde{\rho}(\omega)=\frac{1}{2N}\frac{e^{-\bar{p}|\omega|}+e^{-\bar{q}|\omega|}+e^{-\frac{|\omega|}{2}}-1}
  {1+e^{-|\omega|}}.
\end{eqnarray}
Therefore, the boundary energy can be calculated as
\begin{eqnarray}
     E_{b}(p,q,\xi)
    &=&-2\int_{0}^{\infty}  \frac{e^{-p\omega}}{1+e^{-\omega}} d\omega -2\int_{0}^{\infty}  \frac{e^{-\frac{q}{\sqrt{1+\xi^{2}}}\omega}}
  {1+e^{-\omega}}d\omega \nonumber \\[6pt]
  &&+\pi-2\ln2-1+\frac{1}{p}+\frac{(1+\xi^{2})^{\frac{1}{2}}}{q}.
\label{TL_Eb}
\end{eqnarray}

\begin{figure}[ht]
  \centering
  \includegraphics[width=5.5in]{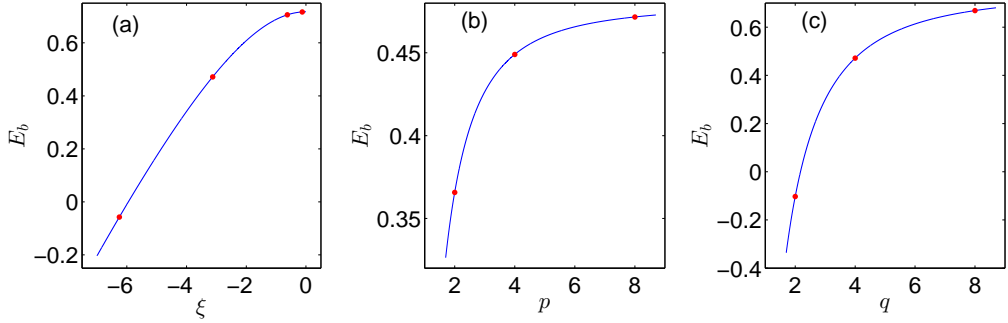}\\
  \caption{The boundary energies versus the boundary parameters. The blue curves are the ones calculated from equation (\ref{TL_Eb}),
  while the red points are the ones obtained from the Hamiltonian (\ref{Hamiltonian}) with the BST algorithms.
Here (a) $p=8$ and $q=4$; (b) $q=4$ and $\xi=-\frac{25}{8}$; (c)
$p=8$ and $\xi=-\frac{25}{8}$. (For interpretation of the references to color in this figure legend, the reader is referred to the web version of this article.)}\label{SU2XXX-Eb_xi_p_q}
\end{figure}

As shown in Figure \ref{SU2XXX-Eb_xi_p_q}, the blue solid lines are the boundary energy calculated by using Eq.(\ref{TL_Eb}),
while the red points are data obtained by employing the BST algorithms \cite{Journal.of.PhysicsA.21.11} to solve the boundary energy of  the Hamiltonian (\ref{Hamiltonian})
in the thermodynamic limit. We can see that the analytical and numerical results agree with each other very well for all tunable parameters.

When $\xi=0$, the non-diagonal boundary condition degenerates into
the diagonal one. The boundary energy (\ref{TL_Eb})
reduce to that of the system with diagonal boundary conditions.
Furthermore, if $\xi$ is small, we can expand  the boundary energy (\ref{TL_Eb}) with respect to $\xi$ as
\begin{eqnarray}\label{Ebxismall}
     E_{b}(p,q,\xi)
    &\simeq& \frac{1}{p}+\psi ^{(0)}\left(\frac{p}{2}\right)-\psi ^{(0)}\left(\frac{p+1}{2}\right)+\frac{1}{q}+\psi ^{(0)}\left(\frac{q}{2}\right)
    -\psi ^{(0)}\left(\frac{q+1}{2}\right) \nonumber \\[6pt]
    &&
    +\pi -1-2 \ln (2)+\xi^2 \left[\frac{1}{2q}-\frac{1}{4}q \psi ^{(1)}\left(\frac{q}{2}\right)+\frac{1}{4}q \psi ^{(1)}\left(\frac{q+1}{2}\right)\right]\nonumber \\[6pt]
    && +\xi^4\frac{\left[q^3 \psi ^{(2)}\left(\frac{q}{2}\right)-q^3 \psi ^{(2)}
    \left(\frac{q+1}{2}\right)+6 q^2 \psi ^{(1)}\left(\frac{q}{2}\right)-6q^2 \psi ^{(1)}\left(\frac{q+1}{2}\right)-4\right]}{32q}\nonumber  \\[6pt]
    && +O\left(\xi^6\right),
\end{eqnarray}
where $\psi ^{(m)}(x)$ is the $m$-order derivative of digamma function \cite{MathematicalSciences17.45}.
The Eq.(\ref{Ebxismall}) contains only even powers of $\xi$ because the energy is invariant under $\xi\rightarrow-\xi$ \cite{JPhysA.47.032001}.
It should be remarked that Eq.(\ref{Ebxismall}) only being effective for small values of $\xi$, while
Eq.(\ref{TL_Eb}) being vailable for general values of $\xi$.

\section{Results for the $su(3)$-invariant spin chain with non-diagonal boundary fields}

\setcounter{equation}{0}

Without losing the generality, we consider a $su(3)$-invariant  spin chain (with the fundamental representation of $su(3)$)  with non-diagonal   boundary fields described by the
Hamiltonian \cite{wang2015off, JournalofHighEnergyPhysics.04.143, Skl88}
\begin{eqnarray}\label{SU3Hamiltonian}
  &&H=2\sum_{j=1}^{N-1}P_{j,j+1}
  +\frac{2\bar{h}}{2+\bar{h}}E_{N}^{13}+\frac{2\bar{h}}{2+\bar{h}}E_{N}^{22}
  +\frac{2\bar{h}}{2+\bar{h}}E_{N}^{31}\nonumber\\
  &&\qquad+ 2hE_{1}^{11}
    +\frac{2}{3}-h, \label{1}
\end{eqnarray}
where the permutation operator is defined in the tensor space of 3-dimensional linear spaces $P_{j,j+1}=\sum_{\mu,\nu=1}^{3}E_{j}^{\mu,\nu}E_{j+1}^{\nu,\mu}$, $E_{j}^{\mu,\nu}$ is the  the Weyl matrix (or the Hubbard operator) $E^{\mu,\nu}=|\mu\rangle\langle\nu|$, $h$ and $\bar{h}$ are arbitrary real
boundary parameters which are related to the boundary fields. The corresponding $K$-matrices $K^{\pm}(u)$ are given  by \footnote{Without losing the generalization, the $K^{\pm}(u)$ given by (\ref{K-3}) and (\ref{K+3}) satisfy  $[K^-(u),K^+(v)]\neq 0$. This fact gives rise to  that  they cannot be diagonalized simultaneously (which corresponds to the non-diagonal (or unparallel) boundary fields), and that there is no an obvious reference state on which the conventional Bethe ansatz \cite{wang2015off} can be performed. }
\begin{eqnarray}
  K^-(u)=1/h+u\left(
     \begin{array}{ccc}
       1 & 0 & 0 \\
       0 & -1 & 0 \\
       0 & 0 & -1 \\
     \end{array}
   \right),\label{K-3}
\end{eqnarray}
and
\begin{eqnarray}
K^+(u)=1/\bar{h}-\left(u+\frac{3}{2}\right)\left(
     \begin{array}{ccc}
       0 & 0 & -1 \\
       0 & -1 & 0 \\
       -1 & 0 & 0 \\
     \end{array}
   \right).\label{K+3}
\end{eqnarray}
The eigenvalue $\Lambda(u)$ of the corresponding transfer matrix is given in terms of the  inhomogeneous $T-Q$
relation \cite{JournalofHighEnergyPhysics.04.143}
\begin{eqnarray}
  \Lambda(u)=z_{1}(u)+z_{2}(u)+z_{3}(u)+x(u),\label{TQ01}
\end{eqnarray}
where the functions $z_{m}(u)$ and $x(u)$
are defined as
\begin{eqnarray}\label{zm}
&&  z_{m}(u) =\frac{u\left(u+\frac{3}{2}\right)K^{(m)}(u)Q^{(0)}(u)}
  {\left(u+\frac{m-1}{2}\right)\left(u+\frac{m}{2}\right)}
  \frac{Q^{(m-1)}(u+1)Q^{(m)}(u-1)}{Q^{(m-1)}(u)Q^{(m)}(u)},\quad
  m=1,2,3, \\[6pt]
\label{x1}
&&    x(u) =u\left(u+\frac{3}{2}\right)Q^{(0)}(u+1)Q^{(0)}(u) \nonumber \\[6pt]
  && \qquad \quad \times\frac{2u\left(u+\frac{1}{2}\right)^{2}\left(u-\frac{1}{2}\right)\left(u+\frac{3}{2}\right)\left(u+1\right)Q^{(2)}(-u-1)}{Q^{(1)}(u)},
\end{eqnarray}
respectively and
\begin{eqnarray}
&&
K^{(1)}(u)=\left(\frac{1}{\bar{h}}+\frac{1}{2}-u\right)\left(\frac{1}{h}+u\right),
\nonumber \\[6pt]
&&
K^{(2)}(u)=\left(\frac{1}{\bar{h}}+\frac{3}{2}+u\right)\left(\frac{1}{h}-u-1\right),\nonumber
  \\[6pt]
&&
K^{(3)}(u)=\left(\frac{1}{\bar{h}}+\frac{3}{2}+u\right)\left(\frac{1}{h}-u-1\right).\label{K03}
\end{eqnarray}
Here the $Q$-functions  can be parameterized as
\begin{eqnarray}
&&  Q^{(0)}(u)=u^{2N},\quad Q^{(3)}=1, \nonumber \\[6pt]
&&
Q^{(r)}(u)=\prod_{l=1}^{L_r}\left(u-\lambda_{l}^{(r)}\right)\left(u+\lambda_{l}^{(r)}+r\right),\quad
r=1, 2, \no \end{eqnarray} where $ L_1=N+L_2+3$ and $0\leq L_2\leq
N$. Then the energy spectrum of the Hamiltonian (\ref{1}) can be
given in terms of the associated Bethe roots by
\begin{eqnarray}\label{SU3E01}
  E=\sum_{l=1}^{L_1}\frac{2}{\lambda_{l}^{(1)}(\lambda_{l}^{(1)}+1)}+2(N-1)
  +\frac{h\bar{h}+2h-2\bar{h}}{2+\bar{h}}+\frac{2}{3},
\end{eqnarray}
where the Bethe roots $\{\lambda^{(r)}_j\}$ should satisfy
the nested BAEs
\begin{eqnarray}
 &&1+\frac{\lambda_l^{(1)}}{\lambda_l^{(1)}+\eta}
 \frac{(2\bar{h}\lambda_l^{(1)}+3\bar{h}+2)(h\lambda_l^{(1)}+h-1)}{(2\bar{h}\lambda_l^{(1)}-\bar{h}-2)(h\lambda_l^{(1)}+1)}
 \frac{Q^{(0)}(\lambda^{(1)}_l+\eta)}{Q^{(0)}(\lambda^{(1)}_l)} \nonumber \\[6pt]
&&\qquad
\times\frac{Q^{(1)}(\lambda^{(1)}_l+\eta)Q^{(2)}(\lambda^{(1)}_l-\eta)}{Q^{(1)}(\lambda^{(1)}_l-\eta)Q^{(2)}(\lambda^{(1)}_l)}
=\bar{c}(\lambda^{(1)}_l)^2\left(\lambda^{(1)}_l+\frac{\eta}{2}\right)^3(\lambda^{(1)}_l+\eta)  \nonumber \\[6pt]
&&\qquad
\times(\lambda^{(1)}_l+\frac{3}{2}\eta)(\lambda^{(1)}_l-\frac{\eta}{2})\frac{Q^{(0)}(\lambda^{(1)}_l)Q^{(2)}(\lambda^{(1)}_l-\eta)}
{Q^{(1)}(\lambda^{(1)}_l-\eta)}, \quad l=1, \cdots, L_1,\label{SU3BAEs1}  \\[6pt]
&&
\frac{(\lambda_k^{(2)}+\frac{3}{2}\eta)}{(\lambda_k^{(2)}+\frac{1}{2}\eta)}
\frac{Q^{(1)}(\lambda^{(2)}_k+\eta)Q^{(2)}(\lambda^{(2)}_k-\eta)}{Q^{(1)}(\lambda^{(2)}_k)Q^{(2)}(\lambda^{(2)}_k+\eta)}=-1,\quad
k=1, \cdots, L_2,\label{SU3BAEs2}
\end{eqnarray}
with
$\bar{c}=4h\bar{h}/[(2\bar{h}\lambda_l^{(1)}-\bar{h}-2)(h\lambda_l^{(1)}+1)]$.

\subsection{Contribution of the inhomogeneous term}

As for the open $su(3)$ quantum spin chain case, the contribution of the inhomogeneous term to the ground state energy $E_{inh}$ is still the same as that defined in the XXX spin-$1/2$ chain
case, i.e.,
\begin{eqnarray}\label{SU3DeltaE}
  E_{inh}=E_{hom}-E_{true},
\end{eqnarray}
where $E_{true}$ is the true values of the ground state energy of the
Hamiltonian (\ref{SU3Hamiltonian}) and $E_{hom}$ is the ground state energy calculated from the energy spectrum
\begin{equation}\label{SU3Energy2}
 E=-\sum_{l=1}^{L_1}\frac{2}{\left(\mu_{l}^{(1)}\right)^{2}+\frac{1}{4}}+2(N-1)
  +\frac{h\bar{h}+2h-2\bar{h}}{2+\bar{h}}+\frac{2}{3},
\end{equation}
where the Bethe roots satisfy the associated BAEs
\begin{eqnarray}
& &\frac{(\mu_l^{(1)}+i\bar{f})(\mu_l^{(1)}+if)}{(\mu_l^{(1)}-i\bar{f})(\mu_l^{(1)}-if)}
 \left(\frac{\mu^{(1)}_l+\frac{i}{2}}{\mu^{(1)}_l-\frac{i}{2}}\right)^{2N+1}
\prod_{j=1}^{L_1}\frac{\left(\mu^{(1)}_l-\mu_{j}^{(1)}-i\right)\left(\mu^{(1)}_l+\mu_{j}^{(1)}-i\right)}
{\left(\mu^{(1)}_l-\mu_{j}^{(1)}+i\right)\left(\mu^{(1)}_l+\mu_{j}^{(1)}+i\right)}
\nonumber \\[6pt]
&&\quad\quad\times\prod_{j=1}^{L_2}\frac{\left(\mu^{(1)}_l+\mu_{j}^{(2)}+\frac{i}{2}\right)\left(\mu^{(1)}_l-\mu_{j}^{(2)}+\frac{i}{2}\right)}
{\left(\mu^{(1)}_l+\mu_{j}^{(2)}-\frac{i}{2}\right)\left(\mu^{(1)}_l-\mu_{j}^{(2)}-\frac{i}{2}\right)}
=-1, \quad l=1, \cdots, L_1, \label{SU3nBAEs01a} \\[6pt]
&&\frac{(\mu_k^{(2)}-\frac{i}{2})}{(\mu_k^{(2)}+\frac{i}{2})}
\prod_{j=1}^{L_1}\frac{\left(\mu^{(2)}_k-\mu_{j}^{(1)}-\frac{i}{2}\right)\left(\mu^{(2)}_k+\mu_{j}^{(1)}-\frac{i}{2}\right)}
{\left(\mu^{(2)}_k-\mu_{j}^{(1)}+\frac{i}{2}\right)\left(\mu^{(2)}_k+\mu_{j}^{(1)}+\frac{i}{2}\right)}
\nonumber \\[6pt]
&&\quad\quad\times\prod_{j=1}^{L_2}\frac{\left(\mu^{(2)}_k-\mu_{j}^{(2)}+i\right)\left(\mu^{(2)}_k+\mu_{j}^{(2)}+i\right)}
{\left(\mu^{(2)}_k-\mu_{j}^{(2)}-i\right)\left(\mu^{(2)}_k+\mu_{j}^{(2)}-i\right)}
=-1,\quad
k=1, \cdots, L_2. \label{SU3nBAEs02a}
\end{eqnarray}
Here we have used the relations $\lambda_l^{(1)}=i\mu_l^{(1)}-\frac{1}{2},
\lambda_k^{(2)}=i\mu_k^{(2)}-1, \bar{f}=-1-1/\bar{h}$ and
$f=-\frac{1}{2}+1/h$.
To simplify the computations we constrain ourselves to the regions\footnote{Similar as the $su(2)$-case discussed in the previous section, it is believed that  the
contribution of the inhomogeneous term to the ground state energy should vanish in the thermodynamic limit for other choices of values of $h$ and $\bar{h}$.
} $h\in(0,2)$, $\bar{h}\in(-1,0)$.
The restrictions imposed on $h$ and $\bar{h}$ are chosen such that $f$ and $\bar{f}$ are positive with
range $(0,\infty)$, for which case  all the Bethe roots of the ground state are real.

Taking the logarithm of BAEs (\ref{SU3nBAEs01a})-(\ref{SU3nBAEs02a}), we obtain
\begin{eqnarray}
&&2\arctan\left(\frac{\mu_l^{(1)}}{\bar{f}}\right)
 +2\arctan\left(\frac{\mu_l^{(1)}}{f}\right)
 +2(2N+1)\arctan\left(2\mu_l^{(1)}\right) \nonumber \\[6pt]
&&\quad\quad-2\sum_{j=1}^{L_1}\arctan\left(\mu^{(1)}_l-\mu_{j}^{(1)}\right)
+\arctan\left(\mu^{(1)}_l+\mu_{j}^{(1)}\right) \nonumber \\[6pt]
&&\quad\quad+2\sum_{j=1}^{L_2}\arctan2\left(\mu^{(1)}_l+\mu_{j}^{(2)}\right)
+\arctan2\left(\mu^{(1)}_l-\mu_{j}^{(2)}\right) =2\pi I_{l}, \no \\[6pt]
&& \qquad\qquad
l=1, \cdots, L_1, \label{SU3nBAEs01b} \\[6pt]
&&2\arctan\left(2\mu_k^{(2)}\right)
+2\sum_{j=1}^{L_1}\arctan2\left(\mu^{(2)}_k-\mu_{j}^{(1)}\right)
+\arctan2\left(\mu^{(2)}_k+\mu_{j}^{(1)}\right) \nonumber \\[6pt]
&&\quad\quad-2\sum_{j=1}^{L_2}\arctan\left(\mu^{(2)}_k-\mu_{j}^{(2)}\right)
+\arctan\left(\mu^{(2)}_k+\mu_{j}^{(2)}\right)=2\pi J_{k}, \no \\[6pt]
&& \qquad\qquad k=1, \cdots, L_2, \label{SU3nBAEs02b}
\end{eqnarray}
where $I_{l}$ and $J_{k}$ are both quantum numbers which determine
the eigenenergy and the corresponding eigenstates.
It is well-known
that the size of the system $N$, with either even or odd value, gives the same
thermodynamic properties. For simplicity, we set $N$ as an even
number and parameterize it as $N=6(n-1)+\alpha$, where $\alpha=2,4,6$.
Then we find that  the values of $L_1$ and $L_2$ in BAEs
(\ref{SU3nBAEs01b})-(\ref{SU3nBAEs02b}) at the ground state are
given by
\begin{eqnarray}\label{SU3L1L2}
    L_1 = L_1^{(\alpha)} +4(n-1),\quad
   L_2 = L_2^{(\alpha)} +2(n-1),
\end{eqnarray}
respectively, where $L_1^{(2)}=2$, $L_2^{(2)}=1$, $L_1^{(4)}=3$, $
L_2^{(4)}=1$, $L_1^{(6)}=4$ and $L_2^{(6)}=2$.

\begin{figure}[ht]
  \centering
  \includegraphics[width=5.5in]{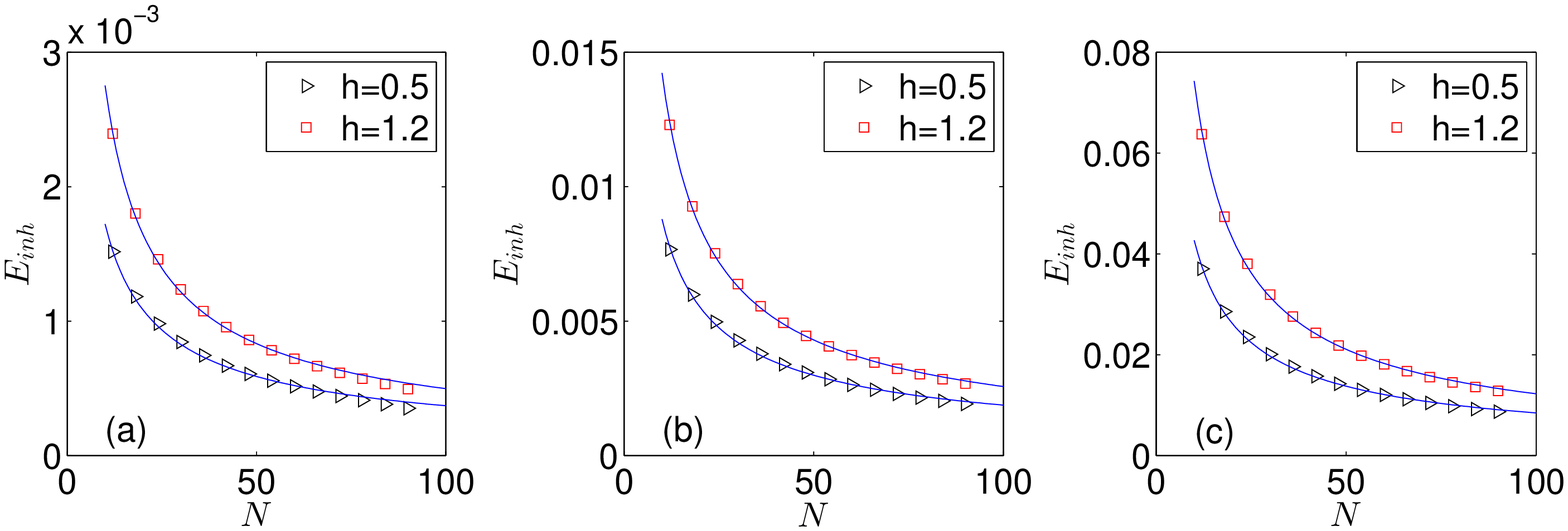}\\
  \caption{Energy $E_{inh}$ as a function of the size of the system $N$.
  The solid lines are the fitting of the numerical data with function $E_{inh}=\gamma_{2}
  N^{\beta_{2}}$. The parameters used are (a) $h=0.5$, $\bar{h}=-\frac{1}{63}$, $\gamma_{2}=0.0080$ and
  $\beta_{2}=-0.6672$; $h=1.2$, $\bar{h}=-\frac{1}{63}$, $\gamma_{2}=0.0152$ and
  $\beta_{2}=-0.7429$; (b) $h=0.5$,  $\bar{h}=-\frac{1}{13}$, $\gamma_{2}=0.0412$ and
  $\beta_{2}=-0.6708$; $h=1.2$,  $\bar{h}=-\frac{1}{13}$, $\gamma_{2}=0.0788$ and
  $\beta_{2}=-0.7434$;
  (c) $h=0.5$,  $\bar{h}=-\frac{1}{3}$,
  $\gamma_{2}=0.2158$ and $\beta_{2}=-0.7035$;
  $h=1.2$,  $\bar{h}=-\frac{1}{3}$,
  $\gamma_{2}=0.4507$ and $\beta_{2}=-0.7829$.}\label{SU3XXXDEvsN}
\end{figure}
To show the finite size effects, we plot the values of
$E_{inh}$ versus the system size $N$ with the choice of
$\bar{h}=-\frac{1}{63}$, $\bar{h}=-\frac{1}{13}$ and
$\bar{h}=-\frac{1}{3}$, while keeping $h=0.5$ and $h=1.2$  in Figure \ref{SU3XXXDEvsN}, respectively. As shown in Figure \ref{SU3XXXDEvsN},
the contribution of the inhomogeneous term to the ground state
energy $E_{inh}$ is a function of the size of the system size $N$ in the form of  $E_{inh}=\gamma_{2}
N^{\beta_{2}}$, where $\beta_{2}<0$ is negative,
which is the same form as the XXX spin-$1/2$ chain case but with different values of the parameters.
In the limiting case where $N$ tends to
infinity, the contribution of $E_{inh}$ to the ground state
energy of the system can be ignored.
 We have checked our results numerically to show that
when $\bar{h}=0$ the system degenerates to the situation of the diagonal
boundary fields, i.e., $E_{inh}=0$.

\subsection{Boundary energy of the open $su(3)$ quantum spin chain}
We have shown in previous section that the contribution of the inhomogeneous term in the
$T-Q$ relation Eq.(\ref{TQ01}) to the ground state energy of the system can be ignored at least in the
thermodynamic limit. This fact allows one to use the homogeneous $T-Q$ relation $\Lambda_{hom}(u)=\sum_{m=1}^{3}z_{m}(u)$,
instead of inhomogeneous one (\ref{TQ01}), to calculate the ground state energy of the system in the thermodynamic
limit by standard method \cite{CambridgeUniversityPress.9780511524332}.
Here, let us consider the case of $f,\bar{f}\geq0$. Following the standard method \cite{CambridgeUniversityPress.9780511524332},
let us  introduce  the so-called the counting functions associated with the two sets of Bethe roots as follows:
\begin{eqnarray}
&&Y(u^{(1)})=\frac{1}{2\pi}\left\{\Xi_{1}\left(u^{(1)}\right)+\frac{1}{2N}\left[\Xi_{1}\left(u^{(1)}\right)
 +\Xi_{2\bar{f}}\left(u^{(1)}\right)
 +\Xi_{2f}\left(u^{(1)}\right)\right]\right\} \nonumber \\[6pt]
&&\qquad\qquad
-\frac{1}{2\pi}\sum_{l=1}^{L_1}\left[\Xi_{2}\left(u^{(1)}-u^{(1)}_l\right)
+\Xi_{2}\left(u^{(1)}+u^{(1)}_l\right)\right] \nonumber \\[6pt]
&&\qquad\qquad
+\frac{1}{2\pi}\sum_{k=1}^{L_2}\left[\Xi_{1}\left(u^{(1)}+u^{(2)}_k\right)
+\Xi_{1}\left(u^{(1)}-u^{(2)}_k\right)\right], \label{SU3nBAEsLimit01} \\[6pt]
&&Z(u^{(2)})=\frac{1}{4N\pi}\left\{\Xi_{1}\left(u^{(2)}\right)\right\}
+\frac{1}{2\pi}\sum_{l=1}^{L_1}\left[\Xi_{1}\left(u^{(2)}-u^{(1)}_l\right)
+\Xi_{1}\left(u^{(2)}+u^{(1)}_l\right)\right] \nonumber \\[6pt]
&&\qquad\qquad
-\frac{1}{2\pi}\sum_{k=1}^{L_2}\left[\Xi_{2}\left(u^{(2)}-u^{(2)}_k\right)
+\Xi_{2}\left(u^{(2)}+u^{(2)}_k\right)\right], \label{SU3nBAEsLimit02}
\end{eqnarray}
with $\Xi_{m}(x)=2\arctan\left(\frac{2x}{m}\right)$. Then the BAEs (\ref{SU3nBAEs01b})-(\ref{SU3nBAEs02b}) become
\begin{equation}
  Y(\mu_{l}^{(1)})=\frac{I_{l}}{2N},\quad   Z(\mu_{k}^{(2)})=\frac{J_{k}}{2N}.\label{BAEs-1}
\end{equation}
In the thermodynamic limit, the Bethe roots (e.g. the solutions to the above equations) for the ground state will become
dense on the real line. This allows one to define the densities of the particles ($\rho(u)$ and $\sigma(u)$) and holes ($\rho^{h}(u)$ and $\sigma^{h}(u)$),
\begin{equation}
\rho(u)+\rho^{h}(u)=\frac{d}{d u}\,Y(u),\quad  \sigma(u)+\sigma^{h}(u)=\frac{d}{d u}\,Z(u).
\end{equation}
Taking the derivative of BAEs (\ref{BAEs-1}) in the thermodynamic limit, we obtain the functional equations of the densities at the ground state\footnote{For the
ground state, the densities of the holes vanish, namely, $\rho^h(u)=\sigma^h(u)=0$.} as
\begin{eqnarray}
&& \rho(u)=a_{1}\left(u\right)+\frac{1}{2N}\left[a_{1}\left(u\right)
 +a_{2\bar{f}}\left(u\right)
 +a_{2f}\left(u\right)-\delta(u)\right] \nonumber \\[6pt]
&&\qquad\qquad
-\int_{-\infty}^{\infty}a_{2}\left(u-\lambda\right)\rho(\lambda)d\lambda
+\int_{-\infty}^{\infty}a_{1}\left(u-v\right)\sigma(v)dv,
\label{nBAES_int01b} \\[6pt]
&&
\sigma(v)=\frac{1}{2N}\left[a_{1}\left(v\right)-\delta\left(v\right)\right]
+\int_{-\infty}^{\infty}a_{1}\left(v-\lambda\right)\rho(\lambda)d\lambda \nonumber \\[6pt]
&&\qquad\qquad
-\int_{-\infty}^{\infty}a_{2}\left(v-\mu\right)\sigma(\mu)d\mu.\label{nBAES_int02b}
\end{eqnarray}

By using Fourier transformation, we obtain $\rho(u)$ as
\begin{eqnarray}
  \rho(u)=\frac{1}{\sqrt{3}(2\cosh(\frac{2\pi u}{3})-1)}.
\end{eqnarray}
Thus the ground state energy takes the form
\begin{eqnarray}\label{E_int01}
  E_{0}(N;h,\bar{h})=-4\pi N\int_{-\infty}^{\infty}a_{1}(\mu)\rho(\mu)d\mu
    +2(N-1)+\frac{h\bar{h}+2h-2\bar{h}}{2+\bar{h}}+\frac{2}{3},
\end{eqnarray}
and the energy density of ground state chooses the value
\begin{eqnarray}
 e_{g}=\lim_{N\rightarrow\infty}\left[\frac{E_{0}(N;h,\bar{h})}{N}\right]=-4\pi
\int_{-\infty}^{\infty}a_{1}(\mu)\rho(\mu)d\mu
    +2+\mathcal{O}(N^{-1}) \simeq-1.406424.
\end{eqnarray}

The boundary energy is given by \cite{PhysRevA.4.386,
JournalofPhysicsA.20.5677, JournalofPhysicsA.23.761}
\begin{eqnarray}
   E_{b}(h,\bar{h})=\lim_{N\rightarrow\infty}\left[E_{0}(N;h,\bar{h})-2E_{0}^{periodic}(N)\right],
\end{eqnarray}
where $E_{0}^{periodic}(N)$ is the ground state energy of the system
with periodic boundary conditions, which can be obtained by the
nested algebraic Bethe ansatz. Using the similar method for the $su(2)$-case in the previous section,
after some tedious calculation, we
obtain the boundary energy for the $su(3)$ spin chain with non-diagonal boundary terms as
\begin{eqnarray}
&&    E_{b}(h,\bar{h})=
   -2\int_{0}^{\infty}
\frac{e^{-\frac{1}{2}\omega-f\omega}+e^{-\frac{3}{2}\omega-f\omega}}
{1+e^{-\omega}+e^{-2\omega}}d\omega \nonumber \\[6pt]
&&\qquad \qquad \quad -2\int_{0}^{\infty}
\frac{e^{-\frac{1}{2}\omega-\bar{f}\omega}+e^{-\frac{3}{2}\omega-\bar{f}\omega}}
{1+e^{-\omega}+e^{-2\omega}}d\omega
+\frac{4\pi}{3\sqrt{3}}+\frac{h\bar{h}+2h-2\bar{h}}{2+\bar{h}}-\frac{4}{3}.
\label{SU3XXXEb}
\end{eqnarray}
\begin{figure}[ht]
  \centering
  \includegraphics[width=5.5in]{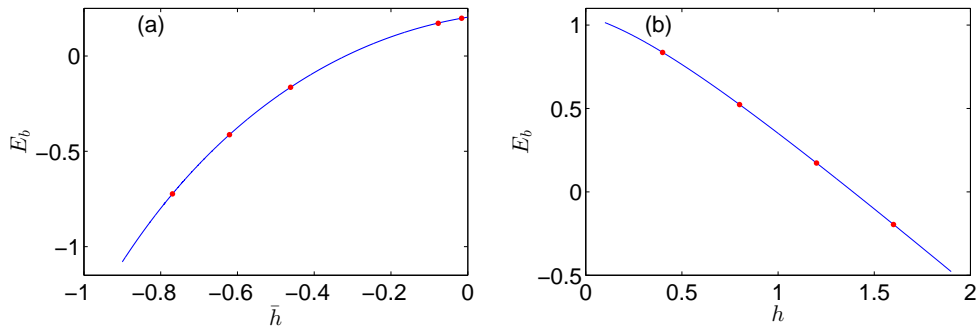}\\
  \caption{Boundary energy as a function of boundary fields. The blue curves are the theoretical results plotted by using equation (\ref{SU3XXXEb}), while the red points are
  the boundary energies obtained by the numerical exact diagonalization and the BST
  extrapolation. Here (a) $h=1.2$; (b) $\bar{h}=-\frac{1}{13}$. (For interpretation of the references to color in this figure legend, the reader is referred to the web version of this article.)}\label{FigSU3XXXEb}
\end{figure}

As shown in Figure \ref{FigSU3XXXEb}, the boundary energies with respect to varying boundary fields $h$ ($\bar{h}$) which are calculated numerically by using exact diagonalization (red dots), where the BST algorithms to obtain the large-N extrapolation of the boundary energy is employed, coincide with the analytical results obtained from Eq.(\ref{SU3XXXEb}) very well.  This also means that the inhomogeneous term in the $T-Q$ relation (\ref{TQ01}) can be neglected in the thermodynamic
limit.

\section{Conclusions}

In this paper, we have studied the thermodynamic limit and boundary
energy of the XXX spin-$1/2$ chain as well as the $su(3)$-invariant spin chain with
unparallel boundary fields. It is shown that the contribution of the
inhomogeneous term in the associated $T-Q$ relation (obtained by the ODBA method)
to the ground state energy does vanish in the thermodynamic limit. This fact allows us to study the
thermodynamics of the model. As concrete examples, we have calculated the boundary energy (\ref{TL_Eb}) for the XXX spin-$1/2$ open chain and
(\ref{SU3XXXEb}) for the $su(3)$ open chain.
The method used in this paper can
be generalized to study the thermodynamic properties and boundary
energy of other high rank models with non-diagonal boundary fields.


\section*{Acknowledgments}
We would like to thank Prof. Y. Wang for his valuable discussions
and continuous encouragement.  The financial supports from the National Natural Science Foundation
of China (Grant Nos. 11375141, 11374334, 11434013, 11425522 and
11547045), the National Program for Basic Research of
MOST (Grant No. 2016YFA0300603), BCMIIS and the Strategic Priority
Research Program of the Chinese Academy of Sciences are gratefully
acknowledged.

\end{document}